\documentstyle[epsf]{mn}

\title{On the spin modulated circular polarization from the
  intermediate polars NY Lup and IGRJ1509-6649.}

\author[Stephen B. Potter et al.]  {Stephen B. Potter$^{1}$\thanks{sbp@saao.ac.za}, Encarni
  Romero--Colmenero$^{1,2}$, Marissa Kotze$^{1,6}$, \and Ewald Zietsman$^{1,3}$,
  O. W. Butters$^{4,7}$, Nikki Pekeur$^{1,5}$ \& David A. H.
  Buckley$^{1,2}$ \\ 
$^{1}$South African Astronomical Observatory, PO Box 9, Observatory 7935, Cape Town, South Africa \\ 
$^{2}$Southern African Large Telescope Foundation, PO Box 9, 7935 Observatory, Cape Town, South Africa \\
$^{3}$Department of Mathematical Sciences, The University of South Africa, PO Box 392, UNISA, 0003, South Africa \\
$^{4}$Department of Physics and Astronomy, University of Leicester, Leicester, LE1 7RH, UK \\
$^{5}$Centre for Space Research, North-West University, Potchefstroom Campus, Private Bag X6001, Potchefstroom 2520, South Africa \\
 $^{6}$Astronomy Department, Astrophysics, Cosmology and Gravity Centre (ACGC), University of Cape Town, Rondebosch 7701, South Africa \\
 $^{7}$Department of Physics and Astronomy, The Open University, Walton Hall,
Milton Keynes, MK7 6AA
}

\date{}

\begin{document}

\maketitle

\begin{abstract}
  
We report on high time resolution, high signal/noise,
photo-polarimetry of the intermediate polars NY Lup and
IGRJ1509-6649. Our observations confirm the detection and colour
dependence of circular polarization from NY Lup and additionally show
a clear white dwarf, spin modulated signal. From our new high
signal/noise photometry we have unambiguously detected wavelength
dependent spin and beat periods and harmonics thereof.  IGRJ1509-6649
is discovered to also have a particularly strong spin modulated
circularly polarized signal. It appears double peaked through the I
filter and single peaked through the B filter, consistent with
cyclotron emission from a white dwarf with a relatively strong
magnetic field.

We discuss the implied accretion geometries in these two systems and
any bearing this may have on the possible relationship with the
connection between polars and soft X-ray-emitting IPs. The relatively
strong magnetic fields is also suggestive of them being polar
progenitors.

\end{abstract}


\section{Introduction}

The standard picture of a cataclysmic variable (CV) is a binary system
consisting of a Roche lobe filling red dwarf (known as the secondary
or the donor star) and an accreting white dwarf (the primary). CVs
have orbital periods of typically a few hours, and mass transfer is
caused by angular momentum loss - see e.g. Warner (1995) for a review
of cataclysmic variables.  Approximately 20\% of the known CVs are
magnetic cataclysmic variables (mCVs), where the white dwarf has a
strong magnetic field (see the catalogue of Ritter \& Kolb
2003). These are further sub-divided into two subtypes, namely
intermediate polars (IPs) and polars, depending on the strength of the
magnetic field of the white dwarf (see e.g. Vrielmann \& Cropper 2004
and Patterson 1994 for a review of these objects).


In IPs it is thought that the white dwarf’s magnetic field truncates
the inner edge of the accretion disc. Accretion is then magnetically
channeled, via accretion curtains, onto the rapidly rotating white
dwarf.  The stronger magnetic field of polars prevents the formation
of a disc entirely and instead accretion occurs onto small localized
region(s), via magnetically confined streams, near the magnetic
pole(s) of the orbitally synchronised white dwarf. Chanmugam \& Ray
(1984) suggested that IPs evolve into polars but, as yet, it is not
fully supported through observations. In particular there is a lack of
polarized emission from IPs compared to polars (in quantity and
magnitude).  It may be possible that the polarized light from IPs is
somehow quenched either by absorption or by emission from the
accretion disc and/or the accretion curtain for example. On the other
hand, as the white dwarfs in IPs rotate asynchronously, it may suggest
that their magnetic moments are less than those in
polars. Wickramasinghe, Wu \& Ferrario (1991) perfored the first
detailed calculation of polarized emission from IPs which showed that
the magnetic fields in IPs is less than 5MG. They also argue that IPs
above the period gap do not evolve into polars below the period gap
but instead the white dwarf remains asynchronous with the resulting
clumpy accretion giveing rise to emission mainly in the
EUV. Additional theoretical studies (Zhang, Wickramasinghe \&
Ferrario 2009) also suggest that if the relatively high mass transfer
rate in IPs, compared to polars, is greater than a critical value,
then the field tends to be advected toward the stellar equator where
it is then buried.

IPs have also been thought of as hard X-ray sources and polars as
softer X-ray sources.  However many IPs in recent X-ray surveys are
shown to have a distinct blackbody component in softer X-rays (e.g.
Mason et al.  1992; Haberl et al. 1994; de Martino et
al. 2004). Haberl \& Motch (1995) suggested that there are two distinct
classes of IP, with the ‘‘soft’’ systems being evolutionary
progenitors of polars.  Evans \& Hellier (2007) made a systematic
analysis of the {\it XMM-Newton} X-ray spectra of IPs and find that
most actually show a soft blackbody component. They put forward that
whether an IP shows a blackbody component depends primarily on
geometrical factors. I.e. in systems that do not show any blackbody
emission is as a result of their heated accretion pole caps being
largely hidden by the accretion curtain. The sample analysed by Evans
\& Hellier additionaly showed that the soft IPs also tended to be polarized
emitters in agreement with the geometrical interpretation.

Therefore good quality multi-filtered polarimetry can, to a first order
approximation, give the magnetic field strength of the white dwarf and
also give insights into the accretion geometry thereby improving our
understanding of these systems. Only eight IPs have been found to emit
polarized light and therefore these new additions represent a
significant increase to the sample. Those IPs where circular
polarization has been found so far are: BG CMi (Penning, Schmidt \&
Liebert 1986; West, Berriman \& Schmidt 1987), PQ Gem (RE 0751+14)
(Rosen, Mittaz \& Hakala 1993; Piirola, Hakala \& Coyne 1993; Potter
et al. 1997), V2400 Oph (RX J1712.6–2414) (Buckley et al. 1997), V405
Aur (RX J0558.0+5353) (Shakhovskoj \& Kolesnikov 1997), V2306 Cyg
(1WGA J1958.2+3232) (Uslenghi et al. 2001; Norton et al. 2002).  1RXS
J173021.5-055933 (Butters et al. 2009), RX J2133.7+5107 (Katajainen et
al. 2007) and NY Lup (Katajainen et al. 2010).



\subsection{Previous observations of NY Lup}

Haberl, Motch \& Zickgraf (2002) identified the {\it ROSAT} source
1RXSJ154814.5-452845 (NY Lup) as an IP using optical and X-ray
observations. They detected the 693 s spin period for the white dwarf
in both the optical and X-ray light curves. NY Lup posseses a highly
absorbed hot black-body soft X-ray component. Furthmore they showed
that the optical spectrum is peculiar in that it shows broad
absorption features underneath Balmer emission lines. de Martino et
al. (2006) obtained the first detailed time resolved optical
spectroscopy from which they determined the orbital period to be 9.87
$\pm$ 0.03 hours, the spectral type of the secondary star to be K2
$\pm$ 2 V, a mass ratio of $q = 0.65 \pm 0.12$, $M_{WD} \ge 0.5
M\sun$, $M_{sec} = 0.4-0.79 M\sun$, a distance of 540-840 pc and a
binary inclination $25^{\rm o} < i \le 58^{\rm o}$. de Martino et
al. (2006) point out that at the distance of NY Lup implies that the
Balmer absorptions are not due to the white dwarf photosphere. Instead
they suggest an orgin in the accretion flow close to the white dwarf
surface. They suggest that polarimetric observations are required to
confirm that NY Lup is a low magnetic field white dwarf. Confirmation
came with Katajainen et al. (2010) who discovered circular
polarization at a level of $\sim$ -1.5\% and $<$0.1\% through the I
and B filters respectively from observations made on the {\it
  VLT}. However, the signal/noise was insufficient to detect
convincingly a modulation on the spin period.

\subsection{Previous observations IGRJ1509-6649}

IGRJ1509-6649 was discovered in the {\it INTEGRAL}/IBIS survey in the
17-60 keV energy range (Revnivtsev et al. 2008) and in the 20-100 keV
energy range (Barlow et al. 2006). Bremsstrahlung and power law fits
gave $kT = 13.8 \pm 5.1$ keV and $\Gamma = 3.6 \pm 0.8$ respectively
with a 20-100 keV flux of $1.38 \times 10^{-11}$ erg s$^{-1}$
cm$^{-2}$. Followup spectroscopy identified it as a probable
intermediate polar (Masetti et al. 2006). Pretorius (2009) published
more detailed followup spectroscopic and photometric observations. The
spectroscopy showed a clear radial velocity signal at 5.89 $\pm$ 0.01
h which was identified as the orbital period. In addition a strong
photometric modulation at 809.42 $\pm 0.02$ s was discovered which was
taken to be the spin period of the magnetic white dwarf. Butters,
Norton, Mukai and Barlow (2009) confirmed the IP classification with
{\it RXTE} observations. The X-ray spin pulse profile is complex with
a modulation depth that decreases with increasing X-ray energy. Their
Bremsstrahlung spectral fit agrees well with (Barlow et al. 2006) with
a column density that suggests absorption within the accretion
flow. They did not find any evidence for an additional modulation at
the beat period and the length of their data set probably precluded
any detection of the orbital period.

\section{Observations and data reduction}
\begin{table*}
\begin{center}
\caption{Table of observations.  All observations were obtained on the
  1.9m telescope using the HIPPO (HI-speed Photo-POlarimeter) of the
  South African Astronomical Observatory.
  {\label{tab:observations}}} \vspace{0.2cm} \centerline{
\begin{tabular}{|l|c|c|c|c|c|c|} \hline
Date             & Target    & Filter      & Total     & length    &Weather    & Brightness\\
                 &           &             & time(hours) & (orbits)  &conditions & \\ 
\hline
20/21 May 2009   & NY Lup    & Clear  & $\sim$2.5   & $\sim$ 0.4& poor & - \\
21/22 May 2009   & IGRJ1509-6649  & Clear  & $\sim$6     & $\sim$ 1.0& good & - \\
22/23 May 2009   & IGRJ1509-6649  & I           & $\sim$6     & $\sim$ 1.0& good & Imag = 14.5-15.3,\\
25/26 May 2009   & NY Lup    & Clear  & $\sim$8     & $\sim$ 0.8& good & - \\
26/27 May 2009   & NY Lup    & B,I         & $\sim$8     & $\sim$ 0.8& good & I,Bmag = 13.6-14,14.6-15\\
24/25 July 2009  & IGRJ1509-6649  & B,I         & $\sim$5     & $\sim$ 0.85& good  & I,Bmag = 14.1-15,15.2-15.6\\
\hline
\end{tabular}
}
\end{center}
\end{table*}

Table ~\ref{tab:observations} shows a log of all the observations of
NY Lup and IGRJ1509-6649. All observations were made with the HIPPO
(HI-speed Photo-POlarimeter: Potter et al. 2010) on the 1.9m telescope
of the South African Astronomical Observatory.  The HIPPO was operated
in its simultaneous linear and circular polarimetry and photometry
mode (All-Stokes). Clear filtered observations (3500-9000\AA) were
defined by the response of the two RCA31034A GaAs photomultiplier
tubes, whilst for others either a B or I filter was used.

Several polarized and non-polarized standard stars (Hsu \& Breger 1982
and Bastien et al. 1988) were observed in order to calculate the
position angle offsets, instrumental polarization and efficiency
factors. Photometric calibrations made use of standard stars chosen
from Landolt (1992). Background sky polarization measurements were
also taken at frequent intervals during the observations. 

Data reduction then proceeded as outlined in Potter et
al. (2010). Figs. ~\ref{NYLup_raw} and ~\ref{IGRJ1509_raw} show a
sample of the reduced and binned observations for NY Lup and
IGRJ1509-6649 respectively. A slightly different data reduction method
was applied in order to produce the spin-phase-fold-binned polarimetry
presented in Figs.~\ref{NYLup_spn_phs_fld} and
~\ref{IGR_spn_phs_fld}. This took advantage of the high speed
capabilities of the HIPPO. Specifically each of the raw, one
millisecond integrations were first phased on the respective spin
periods of the target. Then, for a specified phase bin size, the data
was binned according to the appropriate position of the polarimetry
optics (namely the wave-plate angles). Next, the instrumental
polarization, efficiency factors and sky background were applied to
the binned data. Finally the polarization was calculated for each of
the phase bins. This has the advantage of signifcantly improveing the
S/N of the data by maximizing the amount of signal in each phase and
wave-plate bin before calculating the polarization.

\section{Results}


\subsection{NY Lup}

\begin{figure}
\epsfxsize=8.5cm
\epsffile{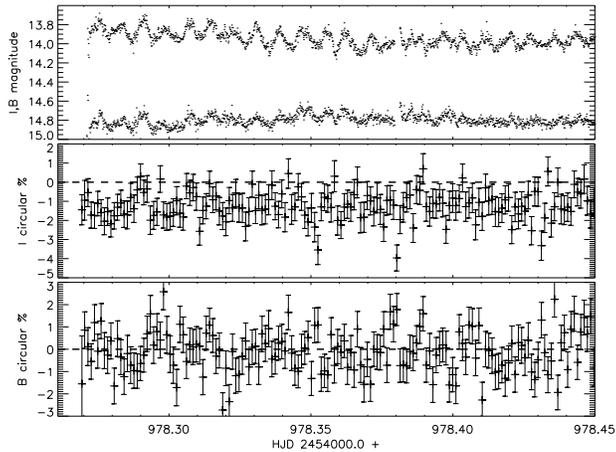} 
\caption{A sample of the simultaneous I and B filtered photometry
  (binned to 10s) and circular polarimetry (binned to 100s) of NY Lup
  taken during the nights of the 26/27 May 2009. The I and B
  photometry are the top and bottom light curves respectively.}
\label{NYLup_raw}
\end{figure}

Fig. ~\ref{NYLup_raw} shows a sample of our observations of NY
Lup. Specifically the simultaneous I and B filtered photo-polarimetric
observations taken on the nights of 26/27 May 2009. The I and B
magnitudes (13.8-14. and 14.7-15.) are broadly consistent with the
observed V magnitudes (14.5-14.7) reported by Haberl et
al. (2002). The 693s white dwarf spin modulation can clearly be picked
out in the I filter photometry. The I filtered circular polarimetry
shows variability mostly confined between -1 and -2 percent with some
excursions to 0 and -3 percent. The B filtered circular polarimetry
also shows variability but appears to be more centered on 0
percent. It is difficult to visually pick out any periodic modulation
in any of the polarimetry given the signal/noise.

We subjected all the photometry and polarimetry to fourier analysis.
In Fig.~\ref{NYLup_amp} we present the amplitude spectra of NY
Lup. The left and right plots display the photometry and corresponding
circular polarimetry respectively. The top plots were constructed from
the 25 May 2009 observations where both HIPPO channels used the clear
filter and were consequently co-added before fourier analysis. The
lower four plots were constructed from the 26 May 2009 observations
where B and I filters were used simultaneously, one in each of the two
channels.

The amplitude spectra of the clear filtered photometry is dominated by
a singular peak centered midway between the known spin and beat
periods (shown as vertical dashed lines) of NY Lup. Ignoring the
lowest frequencies, the I filtered amplitude spectra displays its
largest peak centered on the spin period. The second largest peak is
centered on the beat period. The B filtered amplitude spectra however
does not show any power at the spin period but is instead dominated by
a peak at the beat period.

The corresponding clear filtered circular polarimetry has a
significant peak coincident with the spin period thus formalizing the
discovery of spin modulated circular polarization in NY Lup. A
similarly located peak is present in the I filtered circular
polarimetry, although to a lesser extent. The B filtered circularly
polarized amplitude spectra shows no signal above the noise at the
spin period.

The spin modulation is confirmed in the upper plots of
Fig.~\ref{NYLup_spn_phs_fld} where we have spin-phase-fold-binned the
clear filtered photometry and polarimetry on the spin period of NY Lup
(we have used the NY Lup spin ephemeris of de Martino et
al. 2006). The photometry was normalised by a linear fit before
spin-phase-fold-binning and the error bars represent the standard
deviation of the data in each bin. The photometry and circular
polarimetry show sinusoidal variations roughly in phase with each
other. The circular polarization modulation remains negative
throughout the whole spin cycle between a level of -0.6 and -1.0
percent. Similarly the I filtered photometry is sinusoidally modulated
and in phase with the clear filtered photometry. The I filtered
circular polarimetry is also sinusoidally modulated between
approximately -2 and -2.5 percent, however it is anti-phased with
respect to the photometry and the clear filtered circular
polarimetry. The B filtered photometry shows no modulation on the spin
period, confirming the results of the fourier analysis. However the B
filtered circular polarimetry appears to show a slight sinusoidal
modulation (between approximately 0 and -0.4 percent) which is not too
unexpected given the small peak in the amplitude spectra. It appears
to be in phase with the clear filtered observations.

Linear polarization was detected at a level of $\sim$1.5, $\sim$1.0,
$\sim$2.0 in the clear, I and B filters respectively. No significant
periods were detected in the fourier analysis and no variability is
seen after spin-phase-fold-binning the data (plots not shown).

\begin{figure}
\epsfxsize=8.5cm
\epsffile{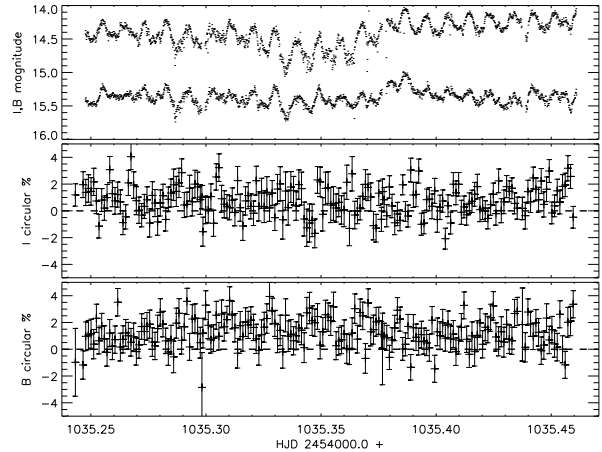} 
\caption{A sample of the simultaneous I and B filtered photometry
  (binned to 10s) and circular polarimetry (binned to 100s) of
  IGRJ1509-6649 taken during the nights of the 24/25 July 2009. The I
  and B photometry are the top and bottom light curves respectively.}
\label{IGRJ1509_raw}
\end{figure}

\begin{figure*}
\epsfxsize=17.5cm
\epsffile{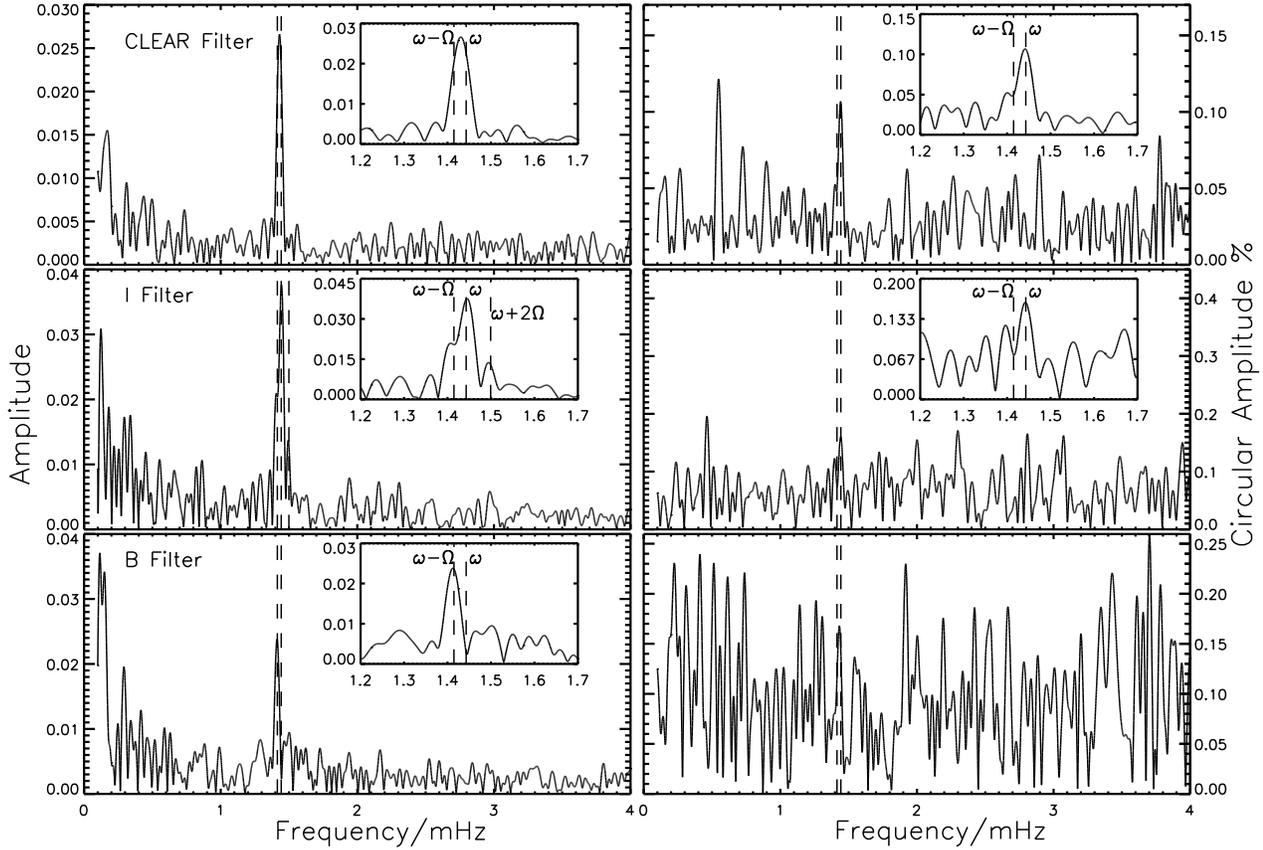} 
\caption{Left and right-hand panels respectively: The photometry and
  circular polarimetry amplitude spectra of NY Lup. The photometry was
  normalised by a linear fit before fourier analysis.}
\label{NYLup_amp}
\end{figure*}

\begin{figure}
\epsfxsize=8.5cm
\epsffile{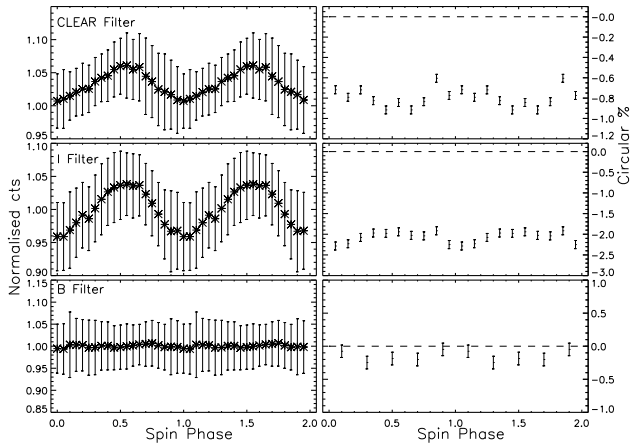} 
\caption{Left and right-hand panels respectively: The
  spin-phase-folded photometry and circular polarimetry of NY Lup. The
  photometry was normalised by a linear fit before phase-folding. The
  I and B filters are simultaneous.}
\label{NYLup_spn_phs_fld}
\end{figure}

\begin{figure*}
\epsfxsize=17.5cm
\epsffile{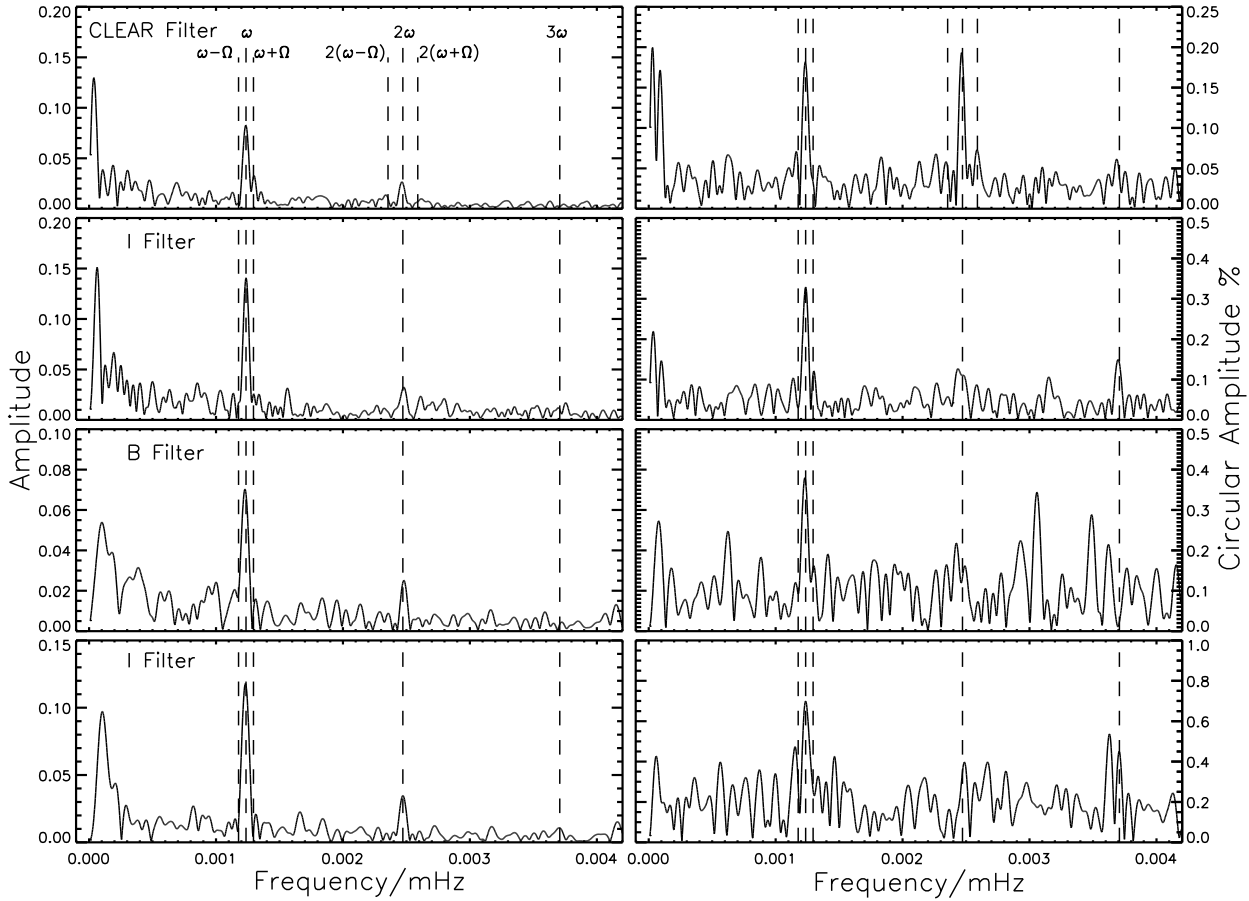} 
\caption{Left and right-hand panels respectively: The photometry and
  circular polarimetry amplitude spectra of IGRJ1509-6649. The
  photometry was normalised by a linear fit before fourier
  analysis. Beat period derived from 1 day alias of Pretorius (2009)
  orbital period.}
\label{IGRJ15094_amp}
\end{figure*}

\begin{figure}
\epsfxsize=8.7cm
\epsffile{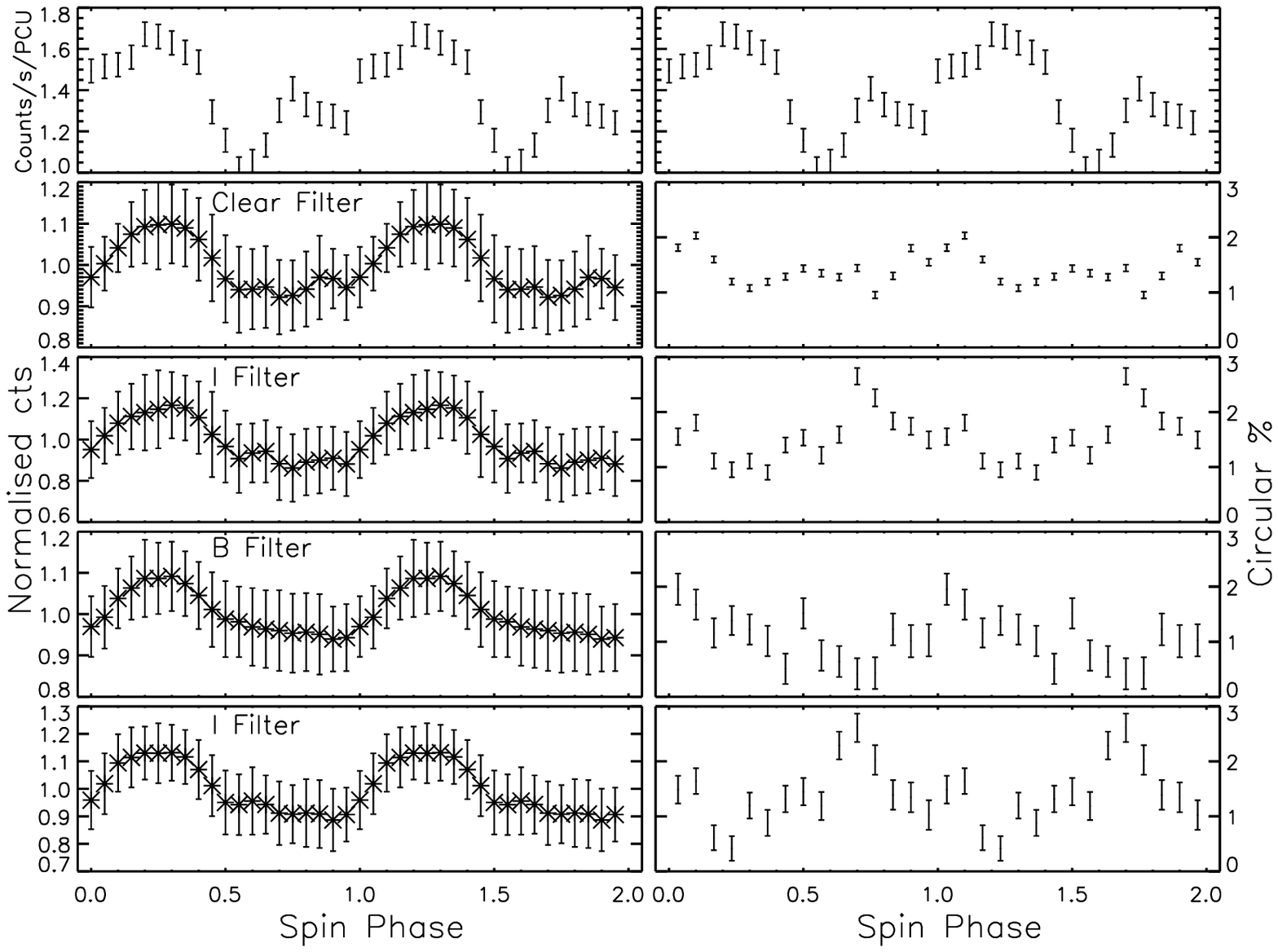} 
\caption{Left and right-hand top panels: RXTE spin-phase-bined
  observations plotted twice. Remaining left and right-hand panels
  respectively: the spin-phase-bined photometry and circular
  polarimetry of IGRJ15094-6649. The photometry was normalised by a
  low order polynomial fit before phase-binning.}
\label{IGR_spn_phs_fld}
\end{figure}

\subsection{IGRJ15094-6649}

Fig. ~\ref{IGRJ1509_raw} shows a sample of our data sets on our
observations of IGRJ15094-6649. Specifically the simultaneous I and B
filtered photo-polarimetric observations taken on the nights of 24/25
July 2009. The I and B magnitudes (15.-14. and 15.7-15.) are broadly
consistent with the observed V magnitudes (15.-14.5) reported by
Pretorius (2009). The 809s white dwarf spin modulation can clearly be
picked out in the I and B filtered photometry. The I and B filtered
circular polarimetry show variability mostly confined between 0 and 2
percent with some excursions to 3 percent.  It is difficult to
visually pick out any periodic modulation in any of the polarimetry
given the signal/noise.

We subjected all the photometry and polarimetry to fourier analysis.
In Fig.~\ref{IGRJ15094_amp} we present the amplitude spectra of
IGRJ15094-6649. The left and right plots display the photometry and
corresponding circular polarimetry respectively. The top plots were
constructed from the 21 May 2009 observations where both HIPPO
channels used the clear filter and were consequently co-added before
fourier analysis. Similarly the next row (of 2 plots) show the results from
co-adding the I filtered observations of 22 May 2009.  The lower four
plots were constructed from the 24 July 2009 observations (1 month
later) where B and I filters were used simultaneously, one in each of
the two channels.

All of the photometric amplitude spectra are dominated by a peak
centered on the known spin period (shown as vertical dashed lines) of
IGRJ15094-6649. A significant peak is also visible at twice the spin
period in all filters. Our data sets are not sufficiently long to
detect the orbital period and the fourier analysis does not show
anything significant at the expected beat period. However, the next
significant peak in the clear filter photometry has a period of
$\sim$773s. If this were the beat period ($\omega + \Omega$) then it
would imply an orbital period of 4.73 hours which would correspond to
a one day alias of the orbital period from Pretorius (2009). Her
figure 2 does indeed show 1 day aliasing. We also calculated the
amplitude spectra of the combined May and July I filtered photometry
(not shown). After pre-whitening the photometry with the spin period,
the next most significant peak (after the spin harmonics) has a period
of 848s. This would be consistent with a ($\omega - \Omega$) beat
period again assuming the orbital period is 4.73 hours. However we
note that this peak is comparable in amplitude to the low frequency
noise and therefore further observations are required to unambigously
identify the orbital period.

The corresponding circular polarimetry (all filters) have a
significant peak coincident with the spin period thus formalizing the
discovery of spin modulated circular polarization in
IGRJ15094-6649. An equally significant peak is located at twice the
spin period in the clear filtered circular polarimetry which may be
present in the 2 sets of I filtered circular polarimetry.

Linear polarization is seen at a level of $\sim$0.5 percent in all
filters (not shown) however the signal/noise is not sufficient to
assign a firm detection. In addition no significant periods were
detected in the fourier analysis.

The spin modulation is confirmed in the left plots of
Fig.~\ref{IGR_spn_phs_fld} where we have spin-phase-fold-binned the
photometry on the spin period of IGRJ15094-6649. The two upper (left
and right) plots show the {\it RXTE} observations from Butters et al
(2009) for comparison (see below).  The photometry was normalised by a
linear fit before spin-phase-fold-binning and the error bars represent
the standard deviation of the data in each bin.  We originally
spin-phase-fold-bined our observations on the spin period of Pretorius
(2009), however the May and July observations showed a phase shift
($\sim 0.15$) with respect to each which we attributed to an
accummulation of the error in the estimate of the white dwarf spin
period. We therefore spin-phase-fold-bined our observations on the
following Barycentric corrected spin ephemeris to bring the data in
line with each other:


$$ T(BJD(tdb)) = 2454973.290 + 0.00936848(4)E $$

BJD is the Barycentric Julian Date in the barycentric dynamic time
system (tdb).  

The photometry shows approximately sinusoidal variations in all
filters. We also extracted and re-reduced the {\it RXTE} observations
orginally published in Butters et al. (2009) and
spin-phase-fold-binned on our new spin ephemeris. Although the {\it
  RXTE} observations show a more complicated variation, the general
morphology appears to be in phase with our new photometry.

The circularly polarized spin-phase-fold-binned data is shown in the
right plots of Fig.~\ref{IGR_spn_phs_fld}. The clear circular
polarimetry varies between $\sim$1.0 and $\sim$2.0 percent, rising
unevenly from its minimum value at phase $\sim$0.3 to a maximum at
phase $\sim$1.1. The uneveness explains the harmonics seen in the
amplitude spectra of Fig.~\ref{IGRJ15094_amp}. The third and fifth
plots show the I filtered circular polarimetry separated by 2
months. They appear to be morphologically identical varying between
$\sim$1.0 and $\sim$2.5 percent with minimum and maximum values at
phases $\sim$0.3 and $\sim$0.7 respectively. The B filtered circular
polarimetry (fourth plot) is simultaneous with the I filtered
observations (fifth plot) and displays a saw tooth variation between
values of $\sim$0.5 percent, at phase $\sim$0.7, and $\sim$2.0 percent
at phase $\sim$0.1. The maximum and minimum between the I and B
filtered circular polarimetry appear to be approximately anti-phased
w.r.t. each other.


\section{Discussion and Summary}

\subsection{NY Lup}

Our photometry confirms the 9.87 h orbital period through the
detection of a strong optical beat period ($\omega - \Omega$) which
particularly dominates in the B filter. It is also present in the I
filter where the spin period is most dominant. A significant ($\omega
+ 2\Omega$) is also detected in the I and B filters. The spin period
is not present in the B filter. The beat period was only weakly
detected in the V filter by Haberl et al. (2002) after demodulation
from the dominant spin frequency. Neither the spin nor the beat
periods were detected in the spectroscopic continuum observations of
de Martino et al. (2006).

Ferrario \& Wickramasinghe (1999) have shown that, for single pole
stream-fed accretion, significant power is expected in the optical at
($\omega - \Omega$). In contrast, for disc accretion, the dominant
power in the continuum and line fluxes is always at the spin frequency
$\omega$. We have found significant power at both ($\omega - \Omega$)
and $\omega$. The single most important feature that allows a clear
distinction to be made between disc-fed and stream-fed accretion is
the amplitude of the radial velocity variations. These have been
measured to be relatively low by de Martino et al. (2006) which
suggests accretion is disc fed. Our observed beat period probably
arises as a result of reprocessing of the primary radiation by other
components of the system such as the accretion curtains, the secondary
star or regions on the accretion disc including the hotspot.

We confirm the detection of circular polarization of Katajainen et
al. (2010) from NY Lup and its B, I colour dependence, although our I
filter observations are higher by $\sim$ 0.5\%. Additionally our
amplitude spectra show unambigously that the circular polarization is
modulated on the white dwarf spin period. The spin modulation was not
detected by Katajainen et al. (2010) probably due to thier lower
signal/noise and/or poorer time resolved {\it VLT} observations. The B
and I filter circular polarization are single humped, negative for the
whole spin cycle, but anti-phased with respect to each other. Buckley
et al. (1995) observed similar behaviour in the IP
RXJ1712.6-2412. Maximum I circular polarization is centered on phase
zero coincident with the observed maximum red shift in the HeII
emission line (de Martino et al 2006). This is consistent with the
orientation of the magnetic field lines, in the cyclotron emission
region, approaching close to parallel to the line of sight at this
phase. This combined with the increase in polarization towards the red
and the colour dependence of the phase for the circular polarization
variation, can be interpreted as an accretion curtain geometry and a
magnetic field $>$ 4MG.  However de Martino et al. (2006) came to the
conclusion that its magnetic field strength is actually lower than 2
MG and Norton et al. (2004) gave estimates $\mu < 1.4 \times 10^{33} G
cm^{3} $ for q = 0.5 for the magnetic moment. Given this disparity it
is not possible to estimate the evolutionary path of NY
Lup. Nethertheless the fact that NY Lup is a ``soft'' IP and shows
spin modulated polarization is consistent with the geometrical
interpretation of IPs (Evans \& Hellier 2007) in order to explain
their X-ray spectral distributions.

We note that the true spectral dependence of the polarimetry is not
yet fully understood because of the strong possibility that some of
the polarized emission in NY Lup originates from halo Zeeman lines (de
Martino 2006). It is thought that halo Zeeman lines are attributed to
free-falling cool material in the vicinity of the shock. This has been
observed in some polars (e.g. MR Ser: Schwope et al. 1993). Further
understanding would require spin-phase resolved spectro-polarimetry
and NY Lup would be a unique opportunity for such an investigation.

Variability in the linear polarization is not detected suggesting that
the magnetic field lines, in the cyclotron emission region, never
approach an orientation close to perpendicular to the line of
sight. This is consistent with a relatively low inclination, single
accreting pole system. The second accretion pole being out of sight.



\subsection{IGRJ15094-6649}




We have found that the IP IGRJ15094-6649 emits circularly polarized
light. The polarization is between $\sim 1.0$ percent and $\sim 2.5$
percent in the I filter and $\sim 1.0$ percent and $\sim 2.0$ percent
in the B filter. 

With an orbital period $>$3 hours and a $P_{spin}/P_{orb} = 0.048$
this puts it with the ``regular'' disc-fed IPs according to Norton,
Wynn \& Somerscales (2004) and Norton et al (2008) and hence it should
share similar properties to the well studied polarized IPs, V2400 Oph
(Buckley et al. 1995) and PQ Gem (Potter et al 1997). The wavelength
and phasing dependence of the positive only circular polarimetry is
almost identical to V2400 Oph, suggesting that it has a single pole-on
geometry with an extended accretion curtain. However the spin period
dominating amplitude spectra (also seen in the X-rays: Butters et al
2009) suggests that it is a disc fed system, more like PQ Gem. In
fact, the X-ray/optical characteristics of IGRJ15094-6649 appear
similar to the disc-fed and accretion curtain scenario of PQ Gem.  In
Fig.~\ref{IGR_spn_phs_fld} we have plotted the {\it RXTE} and our new
observations phased on our new spin ephemeris. As with PQ Gem, maximum
X-ray and optical photometry occur when the accretion region is most
face on to the viewer. Maximum circular polarimetry occurs just before
or after, as one would expect from the angular dependence of cyclotron
beaming.

The detection of significant levels of polarization in the B filter
suggests that the white dwarf's magnetic field could be quite high,
perhaps approaching those of polars ($>10 MG$), but detailed
multi-filtered and/or spectropolarimetric observations are needed to
measure more precisely the wavelength dependence of the circular
polarimetry.  IGRJ15094-6649 would then be similar to the high
magnetic field IP V405 Aurigae (Piirola, Vornanen, Berdyugin and Coyne
2008) and could qualify as a likely candidate as a polar progenitor.

We have tentatively identified a beat period in our clear and I
filtered photometry, which implies an orbital period of 4.73
hours. This corresponds to a one day alias of the orbital period
measured from the radial velocities of the $H_{\alpha}$ line
(Pretorius 2009). This beat period probably arises as a result of
reprocessed radiation on the heated face of the secondary star or a
hot spot on the accretion disc.

Futher observations are required to fully understand this object,
particularly orbit- and spin-resolved spectroscopy and
spectro-polarimetry.


\section{Acknowledgments}
This material is based upon work supported financially by the National
Research Foundation.  Any opinions, findings and conclusions or
recommendations expressed in this material are those of the author(s)
and therefore the NRF does not accept any liability in regard thereto.


\begin{thebibliography}{99}


\bibitem{b2} Barlow E. J., Knigge C., Bird A. J., Dean A. J., Clark
  D. J., Hill A. B., Molina M., Sguera V., 2006, MNRAS, 372, 224


\bibitem {b3} Bastien P., Drissen L., Menard F., Moffat A. F. J.,
  Robert C. \& St-Louis N., 1988, AJ, 95 900




\bibitem{b1} Buckley D. A. H., Haberl F., Motch C., Pollard K.,
  Schwarzenberg-Czerny A. \& Sekiguchi K. 1997, MNRAS, 287, 117



\bibitem{b2} Buckley D. A. H. et al. 1995, MNRAS, 275, 1028



\bibitem{b2} Butters O. W., Norton A. J., Mukai, K. \& Barlow E. J.,
  2009, A\&A, 498, L17

\bibitem{b2} Butters O. W., Katajainen S., Norton A. J., Lehto
  H. J. \& Piirola V., 2009, A\&A. 496, 891



	

\bibitem{b2} Chanmugam G. \& Ray A. 1984, ApJ, 285, 252






\bibitem{b5} de Martino D., Matt G., Belloni T., Haberl F. \& Mukai
  K., 2004, A\&A, 415, 1009

\bibitem{b5} de Martino D., Bonnet-Bidaud J.-M., Mouchet M., Gänsicke
  B. T., Haberl F. \& Motch C., 2006, A\&A, 449, 1151


\bibitem{b5} Evans P. A. \& Hellier C., 2007, ApJ, 663, 1277



  
  
\bibitem{b6} Ferrario L. \& Wickramasinghe D. T., 1999, MNRAS, 309, 517


\bibitem{b6} Haberl F., Motch C. \& Zickgraf F.-J., 2002, A\&A, 387, 201

\bibitem{b6} Haberl F. \& Motch C., 1995, A\&A, 297, 37








\bibitem{b8} Hsu J. C., Breger M., 1982, ApJ, 262, 732



   



\bibitem{b9} Katajainen S., Butters O. W., Norton A. J., Lehto
  H. J. \& Piirola V., 2007, A\&A, 475, 1011

\bibitem{b9} Katajainen S., Butters O., Norton A. J., Lehto H. J.,
  Piirola V. \& Berdyugin A., 2010, ApJ, 724, 165








\bibitem{b8} Landolt, A. U., 1992, AJ, 104, 372

\bibitem{b8} Masetti N., et al. 2006, A\&A, 459, 21

\bibitem{b8} Mason K. O., et al, 1992, MNRAS, 258, 749
















\bibitem{b12} Norton A. J., Quaintrell H., Katajainen S., Lehto H. J.,
  Mukai K. \& Negueruela I., 2002, A\&A, 384, 195

\bibitem{b12} Norton A. J., Wynn G. A. \& Somerscales, 2004, ApJ, 614, 349

\bibitem{b12} Norton A. J., Butters O. W., Parker T. L. \& Wynn G. A., 2008, ApJ, 672, 524





\bibitem{b14} Penning W. R., Schmidt G. D. \& Liebert J., 1986, ApJ, 301, 881

\bibitem{b14} Piirola V., Hakala P. \& Coyne G. V., 1993, AnIPS, 10, 305

\bibitem{b14} Piirola V., Vornanen T., Berdyugin A. \& Coyne G. V., 2008, ApJ, 684, 558

\bibitem{b12} Potter S. B., Cropper M. S., Mason K. O., Hough J. H.,
   Bailey J. A., 1997, MNRAS, 285, 82










\bibitem{b15} Potter S. B., et al. 2010, MNRAS, 402, 1161

\bibitem{b15} Pretorius M. L., 2009, MNRAS, 395, 386



\bibitem{b14} Revnivtsev M., Sazonov S., Krivonos R., Ritter H. \&
  Sunyaev R., 2008, A\&A, 489, 1121



\bibitem{b14} Rosen S. R., Mittaz J. P. D. \& Hakala P. J., 1993, MNRAS, 264, 171










\bibitem{b17} Ritter H., Kolb U., 2003, A\&A, 404, 301 (update RKcat7.9)


\bibitem{b15} Schwope A. D., Beuermann K., Jordan S. \& Thomas, H.-C, 1993, A\&A, 278, 487

  


  


\bibitem{b15} Shakhovskoj N. M. \& Kolesnikov S. V., 1997, IAUC, 6760










\bibitem{b19} Uslenghi M., Tommasi L., Treves A., Piirola V. \& Reig, P., 2001, A\&A, 372, L1



	
\bibitem{b19} Vrielmann S. \& Cropper M, Magnetic Cataclysmic
  Variables, IAU Colloquium 190, 2004, Proceedings of the Conference
  held 8-13 December, 2002 in Cape Town, South Africa. ASP Conference
  Proceedings, Vol. 315. San Francisco: Astronomical Society of the
  Pacific

\bibitem{b22} Warner B. 1995, Cataclysmic Variable Stars, Cambridge
Astrophysics Series 28, Cambridge University Press





\bibitem{b24} West S. C., Berriman G. \& Schmidt G. D., 1987, ApJ, 322, L35

\bibitem{b26} Wickramasinghe D. T., Wu K. \& Ferrario L., 1991, MNRAS, 249, 460

\bibitem{b28} Zhang C. M. Wickramasinghe, D. T. \& Ferrario L. 2009, MNRAS, 397, 2208






\end{thebibliography}
\end{document}